\documentclass[prl,twocolumn,showpacs,superscriptaddress,floatfix]{revtex4}
\usepackage{graphicx}

\newcommand{\CM}{cm}
\newcommand{\upd}{\mbox{\rm d}}

\begin{document}

\title{Strongly Inelastic Granular Gases}
\author{S. Henri Noskowicz} 
\email{henri@eng.tau.ac.il}
\affiliation{Dept. of Fluid Mechanics and Heat Transfer, Tel-Aviv University,
Ramat-Aviv, Tel-Aviv 69978, Israel} 
\author{Oded Bar-Lev} 
\email{fnbarlev@post.tau.ac.il}
\affiliation{Dept. of Fluid Mechanics and Heat Transfer, Tel-Aviv University,
Ramat-Aviv, Tel-Aviv 69978, Israel} 
\author{Dan Serero}
\email{serero@eng.tau.ac.il}
\affiliation{Dept. of Fluid Mechanics and Heat Transfer, Tel-Aviv University,
Ramat-Aviv, Tel-Aviv 69978, Israel} 
\author{Isaac Goldhirsch} 
\email{isaac@eng.tau.ac.il}
\affiliation{Dept. of Fluid Mechanics and Heat Transfer, Tel-Aviv University,
Ramat-Aviv, Tel-Aviv 69978, Israel} 
\date{\today}

\begin{abstract}
The expansion of the velocity distribution
function for the homogeneous cooling state (HCS) in a Sonine polynomial
series around a Maxwellian is shown to be divergent, though Borel resummable. 
A  convergent
expansion for the HCS has been devised and employed  to obtain the HCS velocity
distribution function and (using it) the 
linear transport coefficients for a three dimensional monodisperse
granular gas of smooth inelastic spheres,  for all 
physical values of the coefficient of normal restitution. The results are in
very  good agreement
with findings of  DSMC simulations. 
\end{abstract}

\maketitle

The success of granular hydrodynamics in describing granular gases is by now 
undeniable \cite{goldhirsch03}. However most favorable agreements between
theory and experiment (or simulations) have been restricted to mild degrees
of inelasticity, i.e. coefficients of normal restitution, $\alpha$,  
 that are larger than about 0.7. Furthermore, DSMC simulations seem
to agree with theoretically computed values of the transport coefficients 
\cite{brey05}  for $0.7 \leq \alpha \leq 1$ but definitely
not for $\alpha\leq 0.6$. 
 This is  understandable
for the comparisons with \cite{sela98}, which provides accurate results
for the 
near-elastic case, but less so for 
\cite{brey98} which is formally correct for all values of $\alpha$. 
The reasons for this situation and its resolution are explained below.

Consider a dilute gas of inelastic smooth and homogeneous hard spheres of mass, $m$, diameter, $\sigma$, number density,  $n$,  and coefficient of normal
 restitution, $\alpha$.
 The Boltzmann equation for this case is \cite{sela98,brey98,goldshtein95}:
\begin{equation}
    \frac{\partial f \left( {\bf r},{\bf v}_{1}, t \right) }{\partial t} +
	{\bf v}_{1} {\bf \cdot \nabla} f \left( {\bf r},{\bf v}_{1}, t \right)
	= B \left( f,f\right),
\label{BoltzEq}
\end{equation}
where $f$ denotes the velocity distribution function, ${\bf v}_i$ denotes 
the velocity of a  sphere, 
and ${\bf r}$ the position of its center of mass. 
The  collision term $B(f,f)$, is given by:
\begin{equation}
    B \left( f,f\right) \equiv \sigma^2 \int_{ {\bf k} \cdot {\bf v}_{12} \ >
 \ 0 } \upd \mathbf{v}_{2} \upd
	\mathbf{k} \left( \mathbf{k \cdot v}_{12} \right) \left[
	\frac{1}{\alpha^2} f_1^{\prime} f_2^{\prime} - f_1 f_2 \right],
\label{BolOp}
\end{equation}
where  $f_i \equiv f \left({\bf r}, {\bf v}_{i}, t \right)$, $f_i^{\prime} 
\equiv f
\left({\bf r}, {\bf v}_{i}^{\prime}, t \right)$, $\mathbf{k}$ is a unit vector
along the line of centers of the colliding spheres at contact and primes denote
precollisional velocities. For any set of vectors, $\{ {\bf A}_i \} $:  $\mathbf{A}_{ij} \equiv \mathbf{A}_{i} -
\mathbf{A}_{j}$.  The binary collision between
spheres ``i'' and ``j'' is given by:
\begin{equation}
    \mathbf{v}_{i} = \mathbf{v'}_{i} - \frac{1+\alpha}{2} \left( \mathbf{k}
	\cdot \mathbf{v}'_{ij} \right) \mathbf{k}. \label{KinCol}
\end{equation}
Two systematic perturbative (generalized Chapman-Enskog) expansions 
 for solving Eq.~(\ref{BoltzEq})
 have been developed. Ref.~\cite{sela98}
proposes a double expansion in the gradients (or the Knudsen number,
K) and  the degree of inelasticity $\epsilon \equiv 1-\alpha^{2}$ about a local
Maxwellian, $f_M$. Ref.~\cite{brey98} employs  a gradient expansion
around  the local HCS distribution function,
$f^{ \left(0 \right)}$. Following the  Chapman-Enskog method, it is 
assumed in both
approaches   that $f$
depends on time through the hydrodynamic fields: 
the number density, $n \left({\bf r}, t \right)=\int f
\left({\bf r},{\bf v},t \right)d{\bf v}$, the (granular) 
temperature, $ T \left({\bf r}, t
\right)=\frac{m}{3n}\int f \left({\bf r},{\bf v},t \right)v^2 \upd{\bf v}\equiv
\frac{m}{2}v_{0}^2 \left({\bf r}, t \right)$,   and the velocity field,
 ${\bf u}
\left({\bf r}, t \right)=\int f \left({\bf r},{\bf v},t \right){\bf v} \upd{\bf
v}$. Define the peculiar velocity by:
${\bf V}\equiv {\bf v}-{\bf u}$.  It is convenient to nondimensionalize
 velocities by the thermal
speed  $v_0$; in particular  define: ${\bf c}\equiv {\bf V}/v_0$, and the   
dimensionless distribution function, 
 ${\tilde f}\left({\bf r}, {\bf c}, t
\right) \equiv \frac{v_0^3}{n} f \left({\bf r}, {\bf v}, t \right)$.

Consider first  the HCS. In the absence of gradients, Eq.~(\ref{BoltzEq})
admits a solution, ${\tilde f}^{\left(0
\right)}(c)$, where $c = \left| {\bf c} \right|$ (this `scaling solution'
 is known to be
a `long-time' limit of the homogeneous solution for a non-HCS initial condition,
but this detail is not relevant here). It follows that for the HCS:
\begin{equation}
\label{NormCond}
    \int  {\tilde f}^{\left(0 \right)} \left(c \right)\upd \mathbf{c}=1
	\hspace{0.5cm} ; \hspace{0.5cm}
    \int  c^{2}{\tilde f}^{\left(0 \right)} \left(c \right)\upd
	\mathbf{c}=\frac{3}{2}. 
\end{equation}
Eq.~(\ref{BoltzEq}) becomes 
\begin{equation}
\label{BoltzEqScaled}
    \epsilon W \left(1+ \frac{c_1}{3}\frac{\upd}{\upd c_1} \right) {\tilde
	f}^{\left(0 \right)} = {\tilde B} \left( {\tilde f}^{\left(0 \right)},
	{\tilde f}^{\left(0 \right)}\right), 
\end{equation}
where
\begin{equation}
\label{Wd}
    W  = \frac{\pi}{8} \int \upd \mathbf{c}_{1} \upd	\mathbf{c}_{2} c_{12}^{3}
	{\tilde f}^{\left(0 \right)} \left(c_{1} \right){\tilde f}^{\left(0
	\right)}\left( c_{2} \right), 
\end{equation}
and $\tilde{B} \left( {\tilde f}^{\left(0 \right)},{\tilde f}^{\left(0
\right)} \right)$ is the non-dimensionalized Boltzmann operator
(\ref{BolOp}).
 It is common to represent $\tilde{f}^{(0)}$ in terms of a Sonine polynomial
expansion \cite{goldshtein95,noije98c}: ${\tilde f}^{\left( 0 \right)}\left( c
\right) = {\tilde f}_M \left( c \right) \Phi \left( c\right)$, 
where ${\tilde f}_{M} \left( c
\right) = \pi^{-\frac{3}{2}} e^{-c^{2}}$ is a Maxwellian distribution, and 
 $\Phi \left( c\right) =\sum_{p=0}^{\infty} a_{p}
S_{\frac{1}{2}}^{p} \left( c^{2} \right)$.  The
substitution of this expansion, truncated at $p=2$, in the Boltzmann equation
yields values for the coefficients up to $a_2$, see
\cite{goldshtein95,noije98c}. Ref. \cite{brilliantov06} presents indications
that this expansion diverges and the lack of convergence is
attributed to the well known \cite{esipov97} exponential tail of the HCS (at
large speeds):  ${\tilde f}^{\left( 0 \right)} \left( c \right) \sim
e^{-\frac{3\pi}{\epsilon W} c} \equiv {\tilde f}^{\left(0\right)}_{tail}$ (up
to an algebraic prefactor \cite{hcs}). Indeed, a direct calculation of the
expansion of $\Phi$,  for $\tilde{f}^{\left( 0 \right)}$ replaced by 
$\pi^{\frac{3}{2}}{\tilde f}^{\left( 0 \right)}_{tail}$ (namely the contribution of the tail to the expansion), i.e. an expansion of
$e^{c^{2}}{\tilde f}^{\left( 0 \right)}_{tail}$,  yields \cite{formula}:
 $a_{p} =
\frac{p!}{\sqrt{\pi}} \sum_{q=0}^{p} \frac{\left( - \right)^q \left( q+1
\right)}{\left( p-q \right)!} \left( \frac{2 \epsilon W}{3\pi} \right)^{2q+3}$,
and for large $p$, $a_{p} \sim \left(-\right)^p p \frac{p!}{\sqrt{\pi}} \left(
\frac{2 \epsilon W}{3 \pi} \right)^{2p+3} e^{-\left( \frac{3 \pi}{2 \epsilon
W} \right)^2}$, i.e. the series is divergent, though asymptotic and Borel
resummable. Furthermore, using the method explained below we have computed the
ratio $-\frac{1}{z^2\left(p + 1\right)} \frac{a_{p+1}}{a_p}$ vs. $p$ for the
correct  distribution of the HCS, where $z \equiv \frac{2 \epsilon W}{3 \pi}$,
for $\alpha=0.1$ and $z=0.544$ and find (using $N_s=40$ polynomials, see
Fig.~\ref{Figaps}) that it is essentially constant for $p \geq 5$. It is also
interesting to consider the sum (in which the asymptotic  form of $a_p$ is
substituted for $a_p$): 
\begin{figure}[ht!]
\includegraphics[width=2.5in]{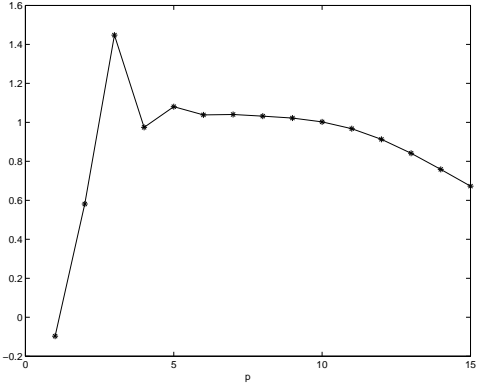}
\caption{$-\frac{1}{z^2\left(p+1\right)} \frac{a_{p+1}}{a_p}$ vs
$p$. $\alpha=0.1$, $z=0.544$ and $N_s=40$. The descent beyond $p=10$ is a result of the truncation error.  \label{Figaps}} 
\end{figure}
$ S\equiv 
    \sum_{p=0}^{\infty}\left(-\right)^p z^{2p} p!S_{\frac{1}{2}}^{p} \left(
    c^{2} \right)$. 
 This sum can be Borel resummed (the Borel sum is denoted by $S_B$) 
upon replacing $p!$
by $\int_{0}^{\infty} e^{-t}t^p \upd t$ and exchanging the orders
of summation and integration:
$ S_B =\int_{0}^{\infty}
e^{-t}\sum_{p=0}^{\infty}
\left(-tz^2\right)^p S_{\frac{1}{2}}^{p} \left( c^{2} \right)\upd t	$.
Next, the sum under the integral can be calculated by inspection noting that
the generating
function of
the Sonine polynomials is  $ \sum_{p=0}^{\infty} s^p S_{m}^{p} \left( x \right) =
\left( 1-s \right)^{-m-1} e^{-x \frac{s}{1-s}}$: 
\begin{equation}
   S_B=  \int_{0}^{\infty} e^{-\left( t +
	\frac{c^2}{1 + z^{2}t} + \frac{3}{2}\ln \left( 1+z^{2}t \right)
	\right)} \upd t.
\label{f0BorelStep2}
\end{equation}
Using the Laplace method one obtains from Eq.~(\ref{f0BorelStep2})
a result which is  practically
identical to $e^{c^2}\tilde{f}^{\left(0\right)}_{tail}$:
 $   S_B \sim 
	\frac{C}{c} e^{-\left(\frac{3\pi}{\epsilon W}c+\frac{\epsilon
	W}{\sqrt{3}\pi c}\right)} $
where $C$ is a constant. Thus we have shown that the standard expansion of
the HCS in Sonine polynomials is divergent, though Borel resummable.

\begin{figure}[ht!]
\includegraphics[width=2.5in]{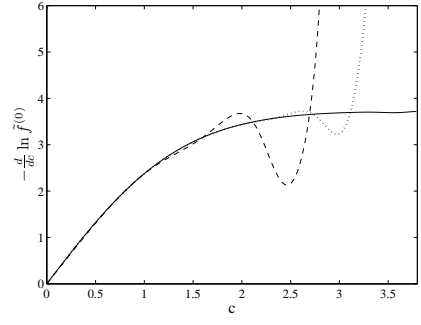}
\caption
{ $-\frac{\upd}{\upd c} \ln \tilde{f}^{\left( 0 \right)}$ vs. $c$ for
$\alpha = 0.1$. The Sonine sum was truncated at $N_S = 10$ (dashes), $N_S = 20$
(dots) and $N_S=40$ (solid  line), to show the convergence towards the 
tail.  \label{FigZanav}}
\end{figure}

 Rather
than using Borel resummation to obtain the distribution function of the
 HCS (and later the linear
transport coefficients) we propose a method that produces   convergent
series for the distribution functions. 
 To this end define the following expansion for the HCS:
\begin{equation}
    {\tilde f}^{\left( 0 \right)} \left( c \right) = \pi^{-\frac{3}{2}}
	e^{-\gamma c^{2}} \sum_{p=0}^{\infty} a^{\left( \gamma \right)}_{p}
	S_{\frac{1}{2}}^{p} \left( c^{2} \right),  
\label{ModifiedSonExp}
\end{equation}
where $ \gamma > 0$ is a constant, i.e. we expand the HCS around a
Maxwellian corresponding to a `wrong' temperature. It can be shown
that 
the coefficients $a^{\left( \gamma \right)}_{p}$ of $e^{c^{2}}{\tilde
f}^{\left(0\right)}_{tail}$ satisfy $\left |a^{\left( \gamma
\right)}_{p}\right | \propto 
\frac{1}{\left(1-\gamma\right)^\frac{3}{2}} \left( \frac{\gamma}{1-\gamma}
\right)^p$,  hence a  convergent expansion is
expected (and obtained) for $\gamma < 1/2$ (in practice
we  chose $\gamma=0.4$).  Upon substituting Eq.~(\ref{ModifiedSonExp}) into
Eqs.~(\ref{NormCond}--\ref{Wd}),  multiplying Eq.~(\ref{BoltzEqScaled}) by
$S_{\frac{1}{2}}^{N} \left( c^{2} \right)$ and integrating over the velocity
one obtains: 
\begin{equation}
\label{NormCondMat}
    \sum_{p=0}^{\infty} Q_p a^{\left( \gamma \right)}_{p} = 1
	\hspace{0.5cm} ; \hspace{0.5cm}
    -\frac{1}{\left( \gamma - 1 \right)^2} \sum_{p=0}^{\infty} Q_p a^{\left(
	\gamma \right)}_{p} p =\frac{3}{2}, 
\end{equation}
\begin{equation}
    \sum_{p,q=0}^{\infty} B_{Npq} a^{\left( \gamma \right)}_{p} a^{\left(
	\gamma \right)}_{q} = \epsilon W \sum_{p=0}^{\infty} M_{Np} a^{\left(
	\gamma \right)}_{p}, \label{BolMat}
\end{equation}
\begin{equation}
\label{WMat}
    W = \sum_{p,q=0}^{\infty} W_{pq} a^{\left( \gamma \right)}_{p} a^{\left(
	\gamma \right)}_{q}, 
\end{equation}
where
\begin{displaymath}
Q_p \equiv \frac{2}{\sqrt{\pi} \gamma^{\frac{3}{2}}} \left(
	\frac{\gamma - 1}{\gamma} \right)^p \frac{\Gamma \left( p +
	\frac{3}{2} \right)}{p!}, 
\end{displaymath}
\begin{displaymath}
M_{Np} \equiv \int \upd \mathbf{c} e^{-\gamma c^2} S_{\frac{1}{2}}^{N}
	\left( c^{2} \right) \left( 1 - \frac{2}{3} \gamma c^2 + \frac{2}{3}
	c^2 \frac{\upd}{\upd c^2} \right) S_{\frac{1}{2}}^{p} \left( c^{2}
	\right),
\end{displaymath}
\begin{displaymath}
W_{pq} \equiv \frac{1}{8 \pi^{\frac{3}{2}}} \int \upd
	\mathbf{c}_{1} \upd \mathbf{c}_{2} e^{-\gamma \left( c_1^{2} + c_2^{2}
	\right)} c_{12}^3 S_{\frac{1}{2}}^{p} \left( c_1^{2} \right)
	S_{\frac{1}{2}}^{q} \left( c_2^{2} \right), 
\end{displaymath}
and 
\begin{widetext}
\begin{equation}\label{bnpq}
{B_{Npq} \equiv \pi^{-3} \int_{{\bf k}\cdot {{\bf c}_{12}}\ >\ 0 } \upd \mathbf{c}_{1} \upd \mathbf{c}_{2} \upd
	\mathbf{k} \left( \mathbf{k \cdot c}_{12} \right) S_{\frac{1}{2}}^{N}
	\left( c_1^{2} \right)}
\left[ \frac{1}{\alpha^2} e^{-\gamma
	\left( c_1^{\prime 2} + c_2^{\prime 2} \right)} S_{\frac{1}{2}}^{p}
	\left( c_1^{\prime 2} \right) S_{\frac{1}{2}}^{q} \left( c_2^{\prime
	2} \right)- e^{-\gamma \left( c_1^{2} + c_2^{2} \right)}
	S_{\frac{1}{2}}^{p} \left( c_1^{2} \right) S_{\frac{1}{2}}^{q} \left(
	c_2^{2} \right) \right]. 
\end{equation}
\end{widetext}
When the above  sums are truncated at $N_S$, one obtains 
a quadratic algebraic system in $N_S + 2$
variables: $W, a^{\left( \gamma \right)}_{0}, \dots , a^{\left( \gamma
\right)}_{N_S}$. The pertinent prefactors are obtained as follows.  Define the  ``super-generating'' function
 \cite{jpc05}:
\[
    I \left(b_{1}, b_{2}, b_{3}, b_{4},x,y,z \right) \equiv 
	\int_{\mathbf{k} \mathbf{\cdot c}_{12} \ > \ 0} \hspace{-0.5cm} \upd \mathbf{c}_{1} \upd
	\mathbf{c}_{2} \upd \mathbf{k} \left( \mathbf{k\cdot c}_{12} \right) e^{-F}, 
\]
where
\begin{eqnarray}
    F & \equiv  & b_{1} c_{1}^{2} + b_{2} c_{2}^{2} + b_{3} c_{1}^{\prime 2} + b_{4}
	c_{2}^{\prime 2} + \frac{x}{2}\left( \mathbf{c}_{1} + \mathbf{c}_{1}^{\prime}
	\right)^2  \nonumber \\
      && + \frac{y}{2} \left( \mathbf{c}_{1} +
	\mathbf{c}_{2}^{\prime} \right)^2 + \frac{z}{2}\left( \mathbf{c}_{1} +
	\mathbf{c}_{2} \right)^2.
    \label{SuperGenDef}
\end{eqnarray}
 This integral can be carried out to yield:
\begin{equation}
    I \left(b_{1}, b_{2}, b_{3}, b_{4},x,y,z \right) = \frac{2
	\pi^{\frac{7}{2}}}{\lambda_1 \lambda_2^{\frac{3}{2}} \left( \lambda_1
	+ \lambda_3 \right)},
    \label{SuperGenSol}
\end{equation}
where $\lambda_1 = \frac{\lambda_2}{4} - \frac{y+z}{2} - \frac{\left(b_{1} -
b_{2} + b_{3} - b_{4} + 2x \right)^2}{4\lambda_2}$, $\lambda_2 =
\sum_{i=1}^{4} b_{i} + 2\left( x+y+z \right)$, $\lambda_3 =
\frac{\epsilon}{4\alpha^2} \left( b_{3} + b_{4} + \frac{x+y}{2} \right) -
\frac{1+\alpha}{4\alpha} \left( x-y \right) - \frac{\lambda_4^2 - 2\lambda_4
\left( b_{1} - b_{2} + b_{3} - b_{4} + 2x \right)}{4\lambda_2}$ and $\lambda_4
= \frac{1 + \alpha}{\alpha} \left( b_{3}-b_{4}+x-y \right)$. 
Next, using Eqs.~(\ref{bnpq},\ \ref{SuperGenSol}) define  a generating function  
from which $B_{Npq}$ can be obtained by taking derivatives (a symbolic processor is used for this purpose):
\begin{eqnarray*}
    \lefteqn{B \left( r, s, t \right) \equiv \sum_{N,p,q=0}^{\infty} r^N s^{p}
	t^{q} B_{Npq} = } && \nonumber \\
      && \pi^{-3} \left( 1-r \right)^{-\frac{3}{2}} \left( 1-s
	\right)^{-\frac{3}{2}} \left( 1-t \right)^{-\frac{3}{2}} \times \nonumber \\
      && \left[ \frac{1}{\alpha^2} I \left( \frac{r}{1-r}, 0, \frac{s}{1-s} +
	\gamma, \frac{t}{1-t} + \gamma, 0, 0, 0 \right) \right. \nonumber \\
      && \left. - I \left( \frac{r}{1-r} + \frac{s}{1-s} + \gamma,
	\frac{t}{1-t} + \gamma, 0, 0, 0, 0, 0, \right) \right]. 
\end{eqnarray*}
The coefficients $W_{pq},\ M_{Np}$ and $Q_p$ are obtained in a similar fashion.
The solution of the $N_S+2$ algebraic equations for $\{ a_n \}$ now enables
one to obtain $\tilde{f}^{\left( 0 \right)}$.  Fig.~\ref{FigZanav} depicts the convergence towards the
exponential tail for  $\alpha=0.1$ through a plot of
 $-\frac{\upd}{\upd c} \ln \tilde{f}^{\left(0\right)}$ vs. $c$ (which should
be constant for an exponential function) for 
$N_S=10,20,40$. The
predicted prefactor in the exponential is $\frac{3 \pi}{\epsilon W}$ and using
Eq.~(\ref{Wd}) it  equals 3.679, while, e.g., for $c=3$,
$-\frac{\upd}{\upd c} \ln \tilde{f}^{\left( 0 \right)} \left( 3 \right) =
3.686$. 

\begin{figure}[ht!]
\includegraphics[width=2.5in]{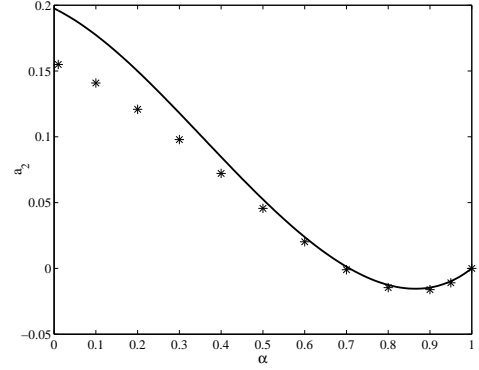}
\caption{A plot of Eq.(\ref{a2Lin}) (solid line) compared to the converged
value of $a_2$ (asterisks)
vs. $\alpha$. \label{Figa2}}
\end{figure}

The coefficient $a_p$ ($\equiv a^{\left(1\right)}_p$) is the projection of
$\tilde{f}^{\left(0\right)}$ on $S_{\frac{1}{2}}^{p}$:
 $   a_p =
\frac{\sqrt{\pi}n!}{2\Gamma\left(n+\frac{3}{2}\right)}\int \upd
	\mathbf{c} S_{\frac{1}{2}}^{p} \left( c^{2} \right) \tilde{f}^{\left(0\right)}\left( c \right)$. Fig.~\ref{Figa2} presents a comparison of the correct
value of $a_2$ with the result obtained from a truncation at $p=2$	
  \cite{goldshtein95,noije98c,brey98}: 
\begin{equation}
    a_2 = \frac{16 \left( 1-\alpha \right) \left( 1-2\alpha^2 \right)}{81 -
	17\alpha + 30 \left( 1-\alpha \right) \alpha^2}. 
\label{a2Lin}
\end{equation}

Next, we apply our results to the theory of Ref.~\cite{brey98}. The 
linear order (in gradients) in the Chapman-Enskog expansion for
Eq.~(\ref{BoltzEq}) is $f^{\left( 1 \right)} = {\bf \cal A} \left( {\bf V} \right)
{\bf \cdot \nabla} \ln T + {\bf \cal B} \left( {\bf V} \right) {\bf \cdot \nabla}
\ln n + {\cal C}_{ij} \left( {\bf V} \right) \overline{\partial_i u_j}$, where an 
overlined tensor  denotes its symmetric and traceless part. The zeroth order cooling rate
$\zeta^{\left( 0 \right)}$ satisfies the following equalities \cite{brey98}:
$\zeta^{\left( 0 \right)} = -\partial^{\left( 0 \right)}_t \ln T = \frac{2
\epsilon}{3} n \sigma^2 v_0 W$ where $W$ is the solution of
Eq.~(\ref{Wd}). Define the linear operator 
\begin{equation}
    Lh \equiv -B \left( f^{\left(0\right)},h\right)-B \left( h,f^{\left(0\right)}\right).
\label{DefL}
\end{equation}
The functions ${\bf \cal A} \left( {\bf V} \right)$, ${\bf \cal B} \left( {\bf V}
\right)$ and ${\cal C}_{ij} \left( {\bf V} \right)$ satisfy \cite{brey98}:
\begin{eqnarray}
    \mathcal{L} \mathbf{\cal A} & = & \frac{\mathbf{V}}{2} \partial_{\mathbf{V}}
	\cdot \left( \mathbf{V} f^{\left( 0 \right)} \right) - \frac{v^2_0}{2}
	\partial_{\mathbf{V}} f^{\left( 0 \right)}, \nonumber \\
    \left( \mathcal{L} + \frac{ \zeta^{\left( 0 \right)}}{2} \right)
	\mathbf{\cal B} & = & \zeta^{\left( 0 \right)} \mathbf{\cal A} - \mathbf{V}
	f^{\left( 0 \right)} - \frac{v^2_0}{2} \partial_{\mathbf{V}} f^{\left(
	0 \right)}, \nonumber \\
    \left( \mathcal{L} + \frac{\zeta^{\left( 0 \right)}}{2}\right)
 {\cal	C}_{ij} &=& \overline{\partial_{V_i}\left(V_jf^{\left(0\right)} \right)},
\label{systf1AEq}
\end{eqnarray}  
where $\mathcal{L} \equiv L - \zeta^{\left( 0 \right)} T \partial_T -
\frac{\zeta^{\left( 0 \right)}}{2}$ and the notation of ref.  \cite{brey98} is
employed. 
 
\begin{figure}[ht!]
\includegraphics[width=2.5in]{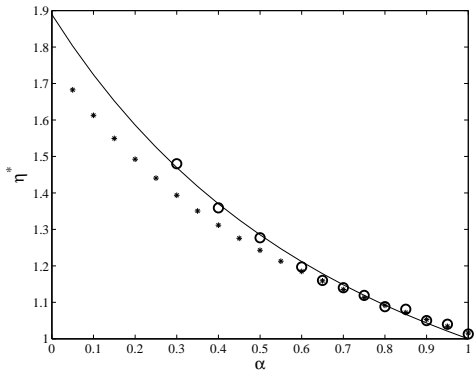}
\caption{The reduced  shear viscosity, $\eta^*$, 
 as a function of
$\alpha$. Asterisks denote the theoretical values and circles 
 represent the results of DSMC simulations. The solid line is the
analytical result obtained in the first Sonine approximation ~\cite{brey98}.   
\label{FigEtaStar}}
\end{figure}

\begin{figure}[ht!]
\includegraphics[width=2.5in]{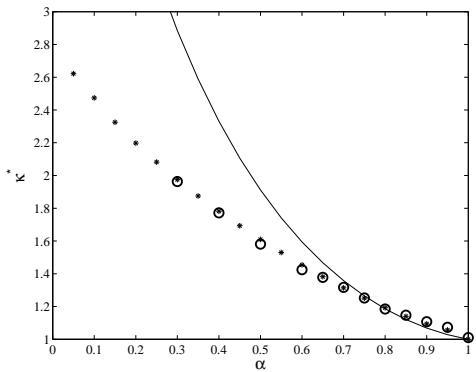}
\caption{The same as in Fig.~(\ref{FigEtaStar}) for $\kappa^*$.
\label{FigKappaStar}}
\end{figure}

\begin{figure}[ht!]
\includegraphics[width=2.5in]{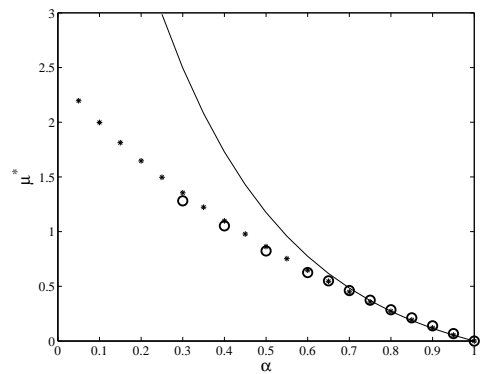}
\caption{ The same as in Fig.~(\ref{FigEtaStar}) for
$\mu^*$.  
\label{FigMuStar}}
\end{figure}

At this order, the pressure tensor, $P_{ij}\equiv m\int \upd\mathbf{v}V_iV_jf$, and the heat flux,
$\mathbf{q}\equiv m/2\int \upd\mathbf{v}V^2 \mathbf{V}f$,  are:
\begin{eqnarray}  
    P^{\left(1\right)}_{ij}&=&m\overline{\partial_ku_l}\int
	\upd\mathbf{v}C_{kl}V_iV_j\equiv-2\eta\overline{\partial_iu_j}, \nonumber \\ 
    q^{\left(1\right)}_i&=&\frac{m}{2}\int \upd\mathbf{v}V^2V_i\left(A_k \partial_k
\ln T+B_k \partial_k\ln n\right) \nonumber \\
&\equiv& -\kappa \  T \ \partial_i\ln T-\mu\  n\  \partial_i\ln n, 
\label{PqSyst}
\end{eqnarray}
thereby defining the shear viscosity,  $\eta$, the thermal
conductivity,   $\kappa$, and the diffusive heat conductivity,  $\mu$. Define
\cite{brey98} the `reduced' coefficients $\eta^*$, $\kappa^*$ 
and $\mu^*$ as the transport 
coefficients normalized by their respective elastic values evaluated
at the lowest order in the Sonine  expansion: 
$\eta^*\equiv \frac{\eta}{\eta_0}$,
$\kappa^*  \equiv 
 \frac{\kappa}{\kappa_0}$, and
$\mu^* \equiv  \frac{n}{T \kappa_0} \mu $, 
where $\eta_0 = \frac{5mv_0}{16\sqrt{2 \pi}\sigma^2}$ and $\kappa_0 = \frac{75
k_Bv_0}{64 \sqrt{2\pi} \sigma^2}$. 
We have solved Eqs.~(\ref{systf1AEq}) using a Sonine polynomial expansion with
$N_S$ up to $10$ (unlike the tail of the
distribution, the transport coefficients depend on low order moments of
$f^{\left( 0 \right)}$), for which convergence was attained.
Figs.~\ref{FigEtaStar},
\ref{FigKappaStar} and \ref{FigMuStar} are plots of $\eta^*$, $\kappa^*$ and
$\mu^*$, respectively as a function of $\alpha$. The agreement between DSMC
simulations (`o') and the  computed coefficients (`*') is excellent, except
in the case of $\eta^*$. For
comparison we have plotted the analytic results obtained in the first Sonine
approximation ~\cite{brey98}. 


In summary, we have shown how a convergent expansion can be successfully
employed to obtain transport coefficients for all physical values of the
coefficient of restitution. The direct  physical relevance of these results 
depends, among other things, on the importance of precollisional correlations 
for strongly dissipative systems (which may require going beyond the 
Boltzmann equation), as well as that of collapse events.  
\begin{acknowledgments}
We are very grateful to Maria Jose Ruiz-Montero for providing us with
the DSMC data used above. We gratefully
  acknowledges partial support from the ISF, grant
  no.\ 689/04, GIF, grant no.\ 795/2003
  and BSF, \mbox{grant no.\ 2004391}.
\end{acknowledgments}
\bibliography{goldhirsch-tgf03}    
\end{document}